\documentclass[twocolumn,showpacs,amsmath,amssymb]{revtex4}
\usepackage{graphicx}
\usepackage{dcolumn}
\usepackage{bm}
\usepackage{epsfig}
\usepackage{amsfonts}

\begin{document}

\title{Recent developments in superstatistics}

\author{Christian Beck}
\affiliation{School of Mathematical Sciences, Queen Mary,
University of London, Mile End Road, London E1 4NS, UK}
\email[E-mail:\ ]{c.beck@qmul.ac.uk}

\date{\today}

\begin{abstract}
We provide an overview on superstatistical techniques applied to
complex systems with time scale separation. Three examples of
recent applications are dealt with in somewhat more detail: the
statistics of small-scale velocity differences in Lagrangian
turbulence experiments, train delay statistics on the British
rail network, and survival statistics of cancer patients once
diagnosed with cancer. These examples correspond to three
different universality classes: Lognormal superstatistics,
$\chi^2$-superstatistics and inverse $\chi^2$ superstatistics.
\end{abstract}

 \maketitle

\section{Introduction}
Many complex systems in various areas of science exhibit a
spatio-temporal dynamics that is inhomogeneous and can be
effectively described by a superposition of several statistics on
different scales, in short a 'superstatistics'
\cite{beck-cohen,unicla,touchette,supergen,souza,chavanis,vignat,Plastino-x,jizba,
wilk,prl01}. The superstatistics notion was introduced in
\cite{beck-cohen}, in the mean time many applications for a
variety of complex systems have been pointed out
\cite{daniels,maya,reynolds,RMT,abul-magd3,porpo,rapisarda,kantz,cosmic,eco,straetennew,eco2}.
Essential for this approach is the existence of sufficient time
scale separation between two relevant dynamics within the complex
system. There is an intensive parameter $\beta$ that fluctuates
on a much larger time scale than the typical relaxation time of
the local dynamics. In a thermodynamic setting, $\beta$ can be
interpreted as a local inverse temperature of the system, but
much broader interpretations are possible.

The stationary distributions of superstatistical systems,
obtained by averaging over all $\beta$, typically exhibit
non-Gaussian behavior with fat tails, which can be a power law, or
a stretched exponential, or other functional forms as well
\cite{touchette}. In general, the superstatistical parameter
$\beta$ need not to be an inverse temperature but can be an
effective parameter in a stochastic differential equation, a
volatility in finance, or just a local variance parameter
extracted from some experimental time series. Many applications
have been recently reported, for example in hydrodynamic
turbulence \cite{prl,beck03,reynolds,unicla}, for defect
turbulence \cite{daniels}, for cosmic rays \cite{cosmic}
and other scattering processes in high energy physics \cite{wilknew, superscatter}, solar
flares \cite{maya}, share price fluctuations
\cite{bouchard,ausloos,eco,straetennew}, random matrix theory
\cite{RMT,abul-magd2,abul-magd3}, random networks
\cite{abe-thurner}, multiplicative-noise stochastic processes
\cite{queiros}, wind velocity fluctuations \cite{rapisarda, kantz},
hydro-climatic fluctuations \cite{porpo}, the statistics of train
departure delays \cite{briggs} and models of the metastatic
cascade in cancerous systems \cite{chen}. On the theoretical
side, there have been recent efforts to formulate maximum entropy
principles for superstatistical systems \cite{souza, abc, crooks,
naudts, straeten}.

In this paper we provide an overview over some recent
developments in the area of superstatistics. Three examples of
recent applications are discussed in somewhat more detail: the
statistics of Lagrangian turbulence, the statistics of train
departure delays, and the survival statistics of cancer patients.
In all cases the superstatistical model predictions are in very
good agreement with real data. We also comment on recent
theoretical approaches to develop generalized maximum entropy
principles for superstatistical systems.

\section{Motivation for superstatistics}\label{sec1}

In generalized versions of statistical mechanics one
starts from more general entropic measures than the
Boltzmann-Gibbs Shannon entropy. A well-known example is the
$q$-entropy \cite{tsallis,mendes}
\begin{displaymath}
S_q =\frac{1}{q-1} (\sum_i p_i^q -1)
\end{displaymath}
but other forms are possible as well (see, e.g., \cite{gentropy}
for a recent review). The $p_i$ are the
probabilities of the microstates $i$, and $q$ is a real number,
the entropic index. The ordinary Shannon entropy is contained as
the special case $q=1$:

\begin{displaymath}
\lim_{q\to 1} S_q =S_1=-\sum_i p_i \log p_i
\end{displaymath}

Extremizing $S_{q}$ subject to suitable constraints yields more
general canonical ensembles, where the probability to observe a
microstate with energy $E_i$ is given by

\begin{displaymath}
p_i \sim e_q^{-\beta E_i}:=\frac{1}{(1+(q-1)\beta
E_i)^\frac{1}{q-1}}
\end{displaymath}

One obtains a kind of power-law Boltzmann factor, of the
so-called $q$-exponential form. The important question is what
could be a physical (non-equilibrium mechanism) to obtain such
distributions.

The reason could indeed be a driven nonequilibrium situation with
local fluctuations of the environment. This is the situation
where the superstatistics concept enters. Our starting point is
the following well-known formula

\begin{displaymath}
\int_0^\infty d \beta f(\beta) e^{-\beta E}
=\frac{1}{(1+(q-1)\beta_0 E)^{1/(q-1)}}
\end{displaymath}

where
\begin{displaymath}
f (\beta) = \frac{1}{\Gamma \left( \frac{1}{q-1} \right)} \left\{
\frac{1}{(q-1)\beta_0}\right\}^{\frac{1}{q-1}}
\beta^{\frac{1}{q-1}-1} \exp\left\{-\frac{\beta}{(q-1)\beta_0}
\right\} \label{fluc}
\end{displaymath}
is the $\chi^2$ (or $\Gamma$) probability distribution.

We see that averaged {\em ordinary} Boltzmann factors with
$\chi^2$ distributed $\beta$ yield a {\em generalized} Boltzmann
factor of $q$-exponential form. The physical interpretation is
that Tsallis' type of statistical mechanics is relevant for {\em
nonequilibrium} systems with temperature fluctuations. This
approach was made popular by two PRLs in 2000/2001
\cite{wilk,prl01}, which used the $\chi^2$-distribution for
$f(\beta)$.
General $f(\beta)$
were then discussed in \cite{beck-cohen}.

In \cite{prl01} it was suggested to construct a {dynamical
realization} of $q$-statistics in terms of e.g. a linear {Langevin equation}

\begin{displaymath}
 \dot{v}=-\gamma v + \sigma L(t)
\end{displaymath}
 with
 fluctuating parameters $\gamma , \sigma$.
Here $L(t)$ denotes Gaussian white noise.
 The parameters
 $\gamma, \sigma$ are supposed to
 fluctuate on a much larger time scale than the 'velocity' $v$.
One can think of a Brownian particle that moves through spatial
'cells' with different local $\beta:=\gamma /(2\sigma^2)$ in each
cell (a nonequilibrium situation).
Assume the probability distribution of $\beta$ in the various
cells is a $\chi^2$-distribution of degree $n$:
\begin{displaymath}
f(\beta) \sim \beta^{n/2-1} e^{-\frac{n\beta}{2\beta_0}}
\end{displaymath}
Then the conditional probability given some fixed $\beta$
in a given cell is Gaussian,
$p(v|\beta))\sim e^{-\frac{1}{2} \beta v^2}$, the joint probability
is $p(v, \beta)=f(\beta) p(v|\beta)$ and the marginal probability
is $p(v)=\int_0^\infty f(\beta) p(v|\beta) d\beta$. Integration
yields
\begin{displaymath}
p(v)\sim \frac{1}{(1+\frac{1}{2}\tilde{\beta}(q-1)v^2)^{1/(q-1)}},
\end{displaymath}
i.e. we obtain power-law Boltzmann factors with
$q=1+\frac{2}{n+1}$, $\tilde{\beta}=2\beta_0/(3-q)$, and
$E=\frac{1}{2}v^2$. Here $\beta_0=\int f(\beta) \beta d\beta $ is
the average of $\beta$.

The idea of superstatistics is to generalize this example to much
broader systems. For example, $\beta$ need not be an inverse
temperature but can in principle be any intensive parameter. Most
importantly, one can generalize to {\em general probability
densities $f(\beta)$} and {\em general Hamiltonians}. In all
cases one obtains a superposition of two different statistics:
that of $\beta$ and that of ordinary statistical mechanics.
Superstatistics hence describes complex {nonequilibrium systems}
with { spatio-temporal fluctuations of an intensive parameter} on
a large scale. The {\em effective} Boltzmann factors $B(E)$ for
such systems are given by

\begin{displaymath}
B(E)=\int_0^\infty f(\beta) e^{-\beta E} d\beta .
\end{displaymath}









Some recent theoretical developments of the superstatistics
concept include the following:

\begin{itemize}

\item Can prove a superstatistical generalization of fluctuation
theorems \cite{supergen}

\item Can develop a variational principle for the large-energy
asymptotics of general superstatistics \cite{touchette}
(depending on $f(\beta)$, one can get not only power laws for
large $E$ but e.g. also stretched exponentials)

\item Can formally define generalized entropies for general
superstatistics \cite{souza,abc, straeten}

\item Can study microcanonical superstatistics (related to a
mixture of $q$-values) \cite{vignat}

\item Can prove a superstatistical version of a Central Limit
Theorem leading to $q$-statistics \cite{Plastino-x}

\item Can relate it to fractional reaction equations \cite{hau}

\item Can consider superstatistical random matrix theory \cite{RMT, abul-magd2, abul-magd3}

\item Can apply superstatistical techniques to networks \cite{abe-thurner}

\item Can define superstatistical path integrals \cite{jizba}

\item Can do superstatistical time series analysis \cite{unicla,kantz, straetennew}
\end{itemize}

...and some more practical applications:

\begin{itemize}

\item Can apply superstatistical methods to analyze the statistics of 3d
hydrodynamic turbulence \cite{beck03,reynolds,unicla,prl}

\item Can apply it to atmospheric turbulence (wind velocity
fluctuations \cite{rapisarda,kantz})

\item Can apply superstatistical methods to finance
and economics \cite{ausloos, eco2, heston, bouchard}

\item Can apply it to blinking quantum dots \cite{grigolini}

\item Can apply it to cosmic ray statistics \cite{cosmic}

\item Can apply it to various scattering processes in particle
physics \cite{wilknew,superscatter}

\item Can apply it to hydroclimatic fluctuations \cite{porpo}

\item Can apply it to train delay statistics \cite{briggs}

\item Can consider medical applications \cite{chen}

\end{itemize}






















\section{Observed universality classes}\label{sec2}

While in principle any $f(\beta)$ is possible in the
superstatistics approach, in practice one usually observes only a
few relevant distributions. These are the $\chi^2$, inverse
$\chi^2$ and lognormal distribution. In other words, in typical
complex systems with time scale separation one usually observes 3
physically relevant universality classes \cite{unicla}
\begin{itemize}
\item (a) $\chi^2$-superstatistics ($=$ Tsallis statistics)
\item (b) inverse $\chi^2$-superstatistics
\item (c) lognormal superstatistics
\end{itemize}

What could be a plausible reason for this? Consider, e.g., case
(a). Assume there are many microscopic random variables $\xi_j$,
$j=1,\ldots , J$, contributing to $\beta$ in an additive way. For
large $J$, the sum $\frac{1}{\sqrt{J}}\sum_{j=1}^J\xi_j$ will
approach a Gaussian random variable $X_1$ due to the (ordinary)
Central Limit Theorem. There can be $n$ Gaussian random variables
$X_1,\ldots ,X_n$ of the same variance due to various relevant
degrees of freedom in the system. $\beta$ should be positive,
hence the simplest way to get such a positive $\beta$ is to
square the Gaussian random variables and sum them up. As a result,
$\beta=\sum_{i=1}^nX_i^2$ is $\chi^2$-distributed with degree $n$,
\begin{equation}
f(\beta )=\frac 1{\Gamma (\frac n2)}\left( \frac n{2\beta
_0}\right) ^{n/2}\beta ^{n/2-1}e^{-\frac{n\beta }{2\beta _0}},
\label{chi2}
\end{equation}
where $\beta_0$ is the average of $\beta$.



(b) The same considerations can be applied if the 'temperature'
$\beta^{-1}$ rather than $\beta$ itself is the sum of several
squared Gaussian random variables arising out of many microscopic
degrees of freedom $\xi_j$. The resulting $f(\beta)$ is the
inverse $\chi^2$-distribution:
\begin{equation}
f(\beta )=\frac{\beta _0}{\Gamma (\frac n2)}\left( \frac{n\beta
_0}2\right) ^{n/2}\beta ^{-n/2-2}e^{-\frac{n\beta _0}{2\beta }}.
\label{chi2inv}
\end{equation}
It generates superstatistical distributions $p(E)\sim \int
f(\beta) e^{-\beta E}$ that decay as $e^{-\tilde{\beta}\sqrt{E}}$
for large $E$.


(c) $\beta$ may be generated by multiplicative random processes.
Consider a local cascade random variable $X_1= \prod_{j=1}^{J}
\xi_j$, where $J$ is the number of cascade steps and the $\xi_j$
are positive microscopic random variables. By the (ordinary)
Central Limit Theorem, for large $J$ the random variable
$\frac{1}{\sqrt{J}} \log X_1= \frac{1}{\sqrt{J}} \sum_{j=1}^J
\log \xi_j$ becomes Gaussian for large $J$. Hence $X_1$ is
log-normally distributed. In general there may be $n$ such
product contributions to $\beta$, i.e., $\beta = \prod_{i=1}^n
X_i$. Then $\log \beta = \sum_{i=1}^n \log X_i$ is a sum of
Gaussian random variables, hence it is Gaussian as well. Thus
$\beta$ is log-normally distributed, i.e.,
\begin{equation}
f(\beta )=\frac{1}{\sqrt{2\pi}s\beta} \exp \left\{ \frac{-(\ln
\frac{\beta}{\mu})^2}{2s^2}\right\}, \label{logno}
\end{equation}
where $\mu$ and $s$ are suitable parameters.


\section{Application to traffic delays}\label{sec3}

We will now discuss examples of the three different
superstatistical universality classes. Our first example is the
departure delay statistics on the British rail network. Clearly,
at the various stations there are sometimes train departure delays
of length $t$. The 0th-order model for the waiting time would be
a Poisson process which predicts that the waiting time
distribution until the train finally departs is $
P(t|\beta)=\beta e^{-\beta t}$, where $\beta$ is some parameter.
But this does not agree with the actually observed data
\cite{briggs}. A much better fit is given by a $q$-exponential,
see Fig.~1.

\begin{figure}
\includegraphics[width=8cm]{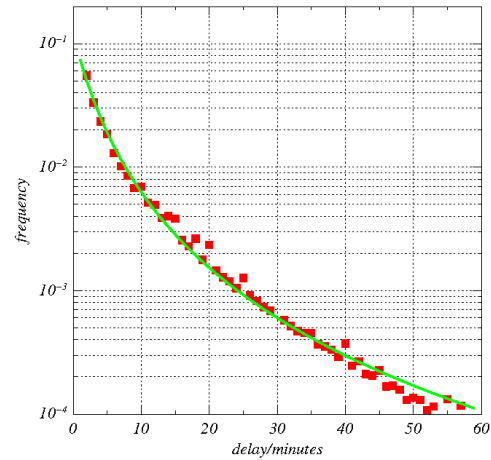}
\caption{Observed departure delay statistics on the British
railway network (data from \cite{briggs}). The solid line is a
q-exponential fit as given by eq.~(\ref{555}).}
\end{figure}

What may cause this power law that fits the data? The idea is that
there are fluctuations in the parameter $\beta$ as well. These
fluctuations describe large-scale temporal or spatial variations
of the British rail network environment. Examples of causes of
these $\beta$-fluctuations:

\begin{itemize}

\item
begin of the holiday season with lots of passengers

\item problem with the track

\item bad weather conditions

\item extreme events such as derailments, industrial action, terror
alerts, etc.

\end{itemize}

As a result, the long-term distribution of train delays is then a
mixture of exponential distributions where the parameter $\beta$
fluctuates:

\begin{equation}
p(t)=\int_0^\infty f(\beta) p(t|\beta) d\beta = \int_0^\infty
f(\beta) \beta e^{-\beta t}. \label{9}
\end{equation}
For a $\chi^2$-distributed $\beta$ with $n$ degrees of freedom
one obtains

\begin{equation}
p(t) \sim {\left( 1+b(q-1)t\right)^{\frac{1}{1-q}}} \label{555}
\end{equation}
where $q=1+{2}/(n+2)$ and $b= {2}\beta_0 /({2-q})$.
The model discussed in \cite{briggs} generates $q$-exponential
distributions of train delays by a simple mechanism, namely a
$\chi^2$-distributed parameter $\beta$ of the local Poisson
process. This is an example for $\chi^2$ superstatistics.

\section{Application to turbulence}\label{sec4}
%



Our next example is an application in turbulence. Consider a
single tracer particle advected by a fully developed turbulent
flow. For a while it will see regions of strong turbulent
activity, then move on to calmer regions, just to continue in yet
another region of strong activity, and so on. This is a
superstatistical dynamics, and in fact superstatistical models
of turbulence have been very successful in recent years
\cite{reynolds,prl}. The typical shape of a trajectory of such a
tracer particle is plotted in Fig.~2.

\begin{figure}
\includegraphics[width=8cm]{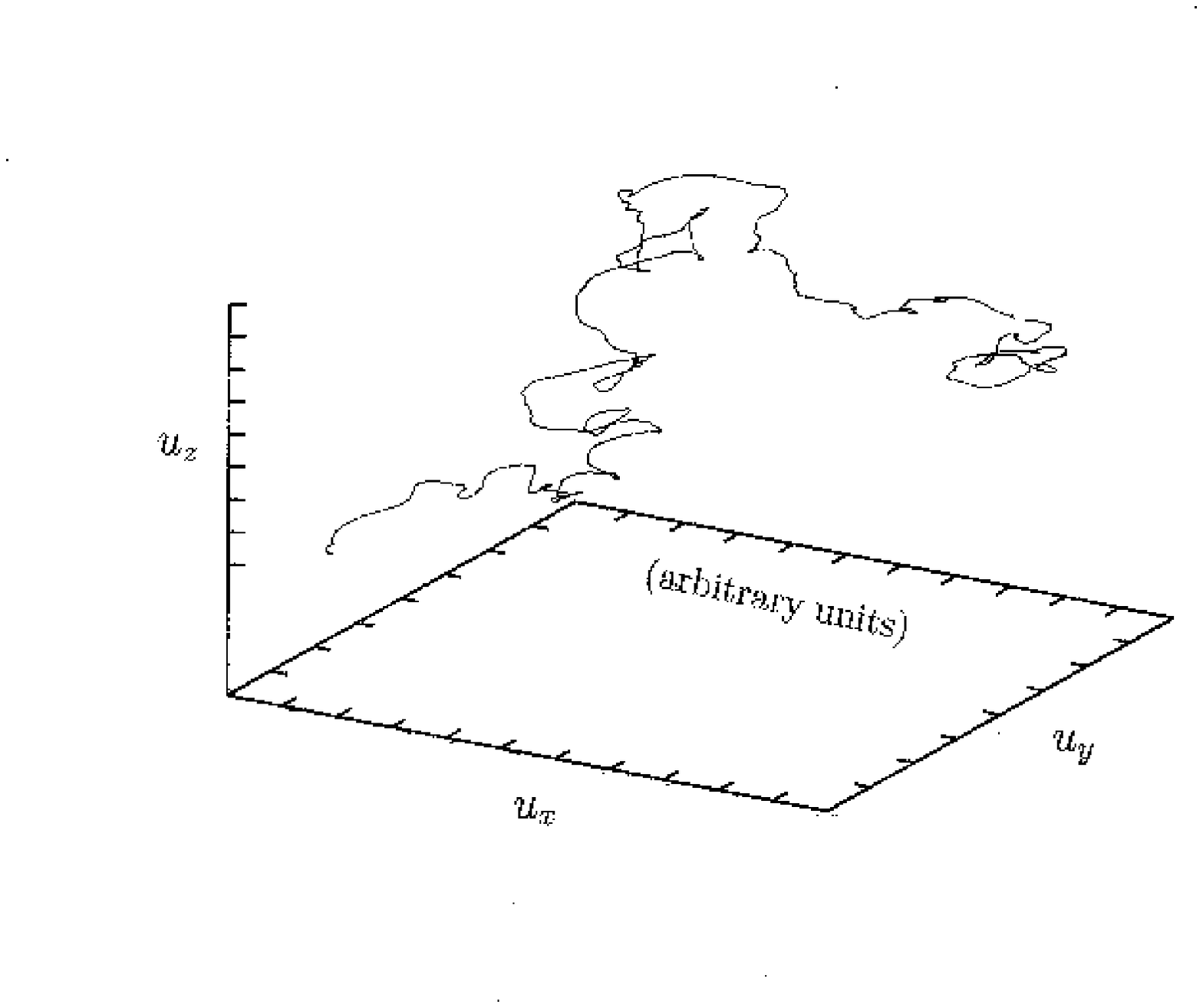}
\caption{Typical example of a trajectory of a test particle in a
turbulent flow.}
\end{figure}

This is 'Lagrangian turbulence' in contrast to 'Eulerian
turbulence', meaning that one is following a single particle in
the flow. In particular, one is interested in velocity differences
$\vec{u}(t):=\vec{v}(t+\tau) -\vec{v} (t)$ of the particle on a
small time scale $\tau$. For $\tau \to 0$ this velocity
difference becomes the local acceleration $\vec{a}
(t)=\vec{u}(t)/\tau$. A superstatistical Lagrangian model for
{3-dim} velocity differences of the tracer particle has been
developed in \cite{prl}. One simply looks at a superstatistical
Langevin equation of the form

\begin{equation}
\dot{\vec{u}}=-\gamma \vec{u}+B \vec{n} \times \vec{u}+\sigma
\vec{L}(t). \label{2}
\end{equation}
Here $\gamma$ and $B$ are constants. Note that the term
proportional to $B$ introduces some rotational movement of the
particle, mimicking the vortices in the flow. The noise strength
$\sigma$  and the unit vector $\vec{n}$ evolve stochastically on
a large time scale $T_\sigma$ and $T_{\vec{n}}$, respectively,
thus obtaining a superstatistical dynamics. $T_\sigma$ is of the
same order of magnitude as the integral time scale $T_L$, whereas
$\gamma^{-1}$ is of the same order of magnitude as the Kolmogorov
time scale $\tau_{\eta}$. One can show that the Reynolds number
$R_\lambda$ is basically given by the time  scale ratio $T_\sigma
\gamma \sim T_L/\tau_{\eta} \sim R_\lambda
>>1$.
The
time scale $T_{\vec{n}}>> \tau_\eta$ describes the average life
time of a region of given vorticity surrounding the test particle.

In this superstatistical turbulence model one defines the parameter
$\beta$ to be
$\beta:=2\gamma/\sigma^2$, but it does {\em not} have the meaning
of a physical inverse temperature in the flow. Rather, one has
$ \beta^{-1} \sim \nu^{1/2} \langle \epsilon \rangle^{-1/2}
\epsilon$, where $\nu$ is the kinematic viscosity and $\langle
\epsilon \rangle$ is the average energy dissipation, which is known to fluctuate in
turbulent flows. In fact, Kolmogorov's theory of 1961 suggests a lognormal
distribution for $\epsilon$, which automatically leads us to lognormal
superstatistics: It is reasonable to assume that the
probability density of the stochastic process $\beta(t)$ is
approximately a lognormal distribution
\begin{equation}
f(\beta) = \frac{1}{\beta s \sqrt{2\pi}}\exp\left\{ \frac{-(\log
\frac{\beta}{m})^2}{2s^2}\right\}. \label{Logno}
\end{equation}
For very small $\tau$ the 1d acceleration component of the
particle is given by $a_x=u_x/\tau$ and one gets out of the model
the 1-point distribution
\begin{equation}
p(a_x) = \frac{\tau}{2\pi s }\int_0^\infty d\beta \; \beta^{-1/2}
\exp\left\{ \frac{-(\log \frac{\beta}{m})^2}{2s^2}\right\}
e^{-\frac{1}{2}\beta \tau^2 a_x^2} . \label{10}
\end{equation}
This prediction agrees very well with experimentally measured
data of the acceleration statistics, which exhibits very
pronounced (non-Gaussian) tails, see Fig.~3 for an example.

\begin{figure}
\includegraphics[width=8cm]{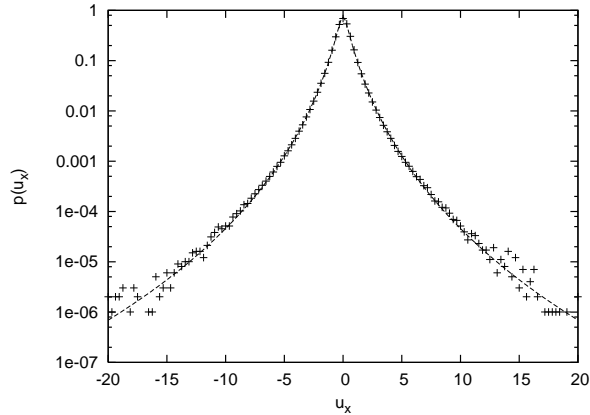}
\caption{Measured probability density of small-scale velocity
differences in Lagrangian turbulence (data from \cite{pinton}).
The solid line is a fit of the form (\ref{10}) (see \cite{prl}
for more details).}
\end{figure}

It is interesting to see that our 3-dimensional superstatistical
model predicts the existence of correlations between the
acceleration components. For example, the acceleration $a_x$ in
$x$ direction is not statistically independent of the
acceleration $a_y$ in $y$-direction. We may study the ratio
$R:=p(a_x,a_y)/(p(a_x)p(a_y))$ of the joint probability
$p(a_x,a_y)$ to the 1-point probabilities $p(a_x)$ and $p(a_y)$.
For independent acceleration components this ratio would always
be given by $R=1$. However, our 3-dimensional superstatistical
model yields prediction
\begin{equation}
R=\frac{\int_0^\infty \beta f(\beta)e^{-\frac{1}{2}\beta \tau^2
(a_x^2+a_y^2)}d\beta}{ \int_0^\infty\beta^{1/2}f(\beta
)e^{-\frac{1}{2}\beta \tau^2 a_x^2}d\beta
\int_0^\infty\beta^{1/2}f(\beta)e^{-\frac{1}{2}\beta \tau^2
a_y^2}d\beta} \label{999}
\end{equation}
This is a very general formula, valid for any superstatistics,
for example also Tsallis statistics, obtained when $f(\beta)$ is
the $\chi^2$-distribution. The trivial result $R=1$ is obtained
only for $f(\beta)=\delta (\beta -\beta_0)$, i.e.\ no
fluctuations in the parameter $\beta$ at all. Fig.~4 shows
$R:=p(a_x,a_y)/(p(a_x)p(a_y))$ as predicted by lognormal
superstatistics:
\begin{figure}
\includegraphics[width=8cm]{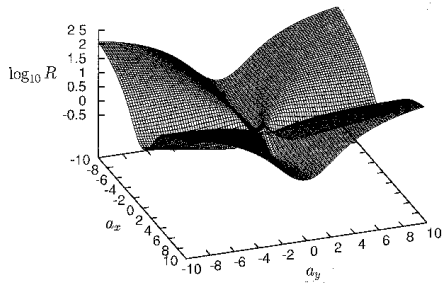}
\caption{Predicted shape of correlations as given by
eq.~(\ref{999}), $f(\beta)$ being the lognormal distribution.}
\end{figure}
The shape of this is very similar to experimental measurements
\cite{boden-reynolds,prl}.

\section{Application in medicine}\label{sec5}

Our final example of application of superstatistics is for a
completely different area: medicine. We will look at cell
migration processes describing the metastatic cascade of
cancerous cells in the body \cite{chen}.
There are various pathways in which cancerous cells can migrate:
Via the blood system, the lymphatic system, and so on. The
diffusion constants for these various pathways are different. In
this way superstatistics enters, describing different diffusion
speeds for different pathways (see Fig.~5).
\begin{figure}
\includegraphics[width=8cm]{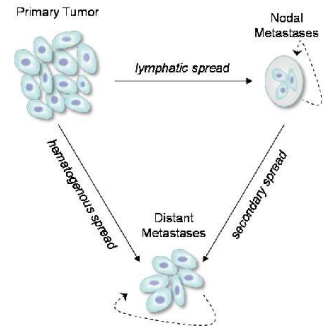}
\caption{Various pathways for the spread of cancer cells}
\end{figure}
But there is another important issue here: When looking at a large
ensemble of patients then the spread of cancerous cells can be very different
from patient to patient. For some patients the
cancer spreads in a very aggressive way, whereas for others it is much slower and less aggressive.
So superstatistics also arises from the fact that all patients are different.

A superstatistical model of metastasis and cancer survival has
been developed in \cite{chen}. Details are described in that
paper. Here we just mention the final result that comes out of the
model: One obtains the following prediction for the probability
density function of survival time $t$ of a patient that is
diagnosed with cancer at $t=0$:
\begin{equation}
p \left( t \right) = \int _{0}^{\infty }\!{\frac
{{t}^{n-1}{\lambda}^{n}{e^{-\lambda\,t}}}{ \Gamma \left( n
\right)} \frac{\lambda_0{\left(n\lambda_0 /
2\right)}^{n/2}}{\Gamma\left(n / 2\right)} \lambda^{-n/2-2}
e^{\frac{-n\lambda_0}{2\lambda}}}{d\lambda}, \label{eq6}
\end{equation}
or
\begin{eqnarray}
p \left( t \right) &=&
\frac{(n\lambda_0)^{3n/4}}{\Gamma(n)\Gamma(n/2)} \left(
\frac{t}{2} \right)^{3n/4-1} \left[ \frac{\sqrt{2n\lambda_0t}}{n}
K_{n/2+1}\left( \sqrt{2n\lambda_0t} \right) \right. \nonumber \\
 &-& \left. K_{n/2}\left(
\sqrt{2n\lambda_0t} \right) \right], \label{eq7}
\end{eqnarray}
where $K_\nu(z)$ is the modified Bessel function. Note that this
is inverse $\chi^2$ superstatistics. The role of the parameter
$\beta$ is now played by the parameter $\lambda$, which in a sense
describes how aggressively the cancer propagates.

The above formula based on inverse $\chi^2$ superstatistics is in
good agreement with real data of the survival statistics of
breast cancer patients in the US. The superstatistical formula
fits the observed distribution very well, both in a linear and
logarithmic plot (see Fig.6).

One remark is at order. When looking at the relevant time scales one should
keep in mind
that the data shown are survival distributions {\em conditioned on the fact that death occurs
due to cancer}. Many patients, in particular if they are diagnosed at an early stage, will live a long happy life
and die from something else than cancer. These cases are {\em not} included in the
data.

\begin{figure}
\includegraphics[width=8cm]{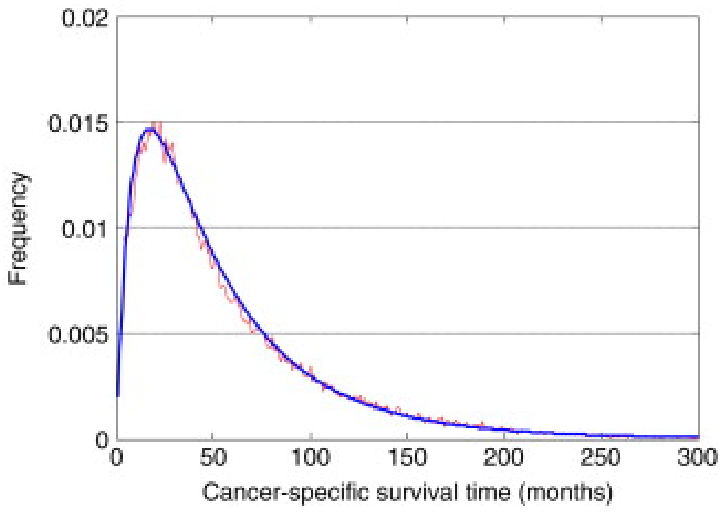}
\includegraphics[width=8cm]{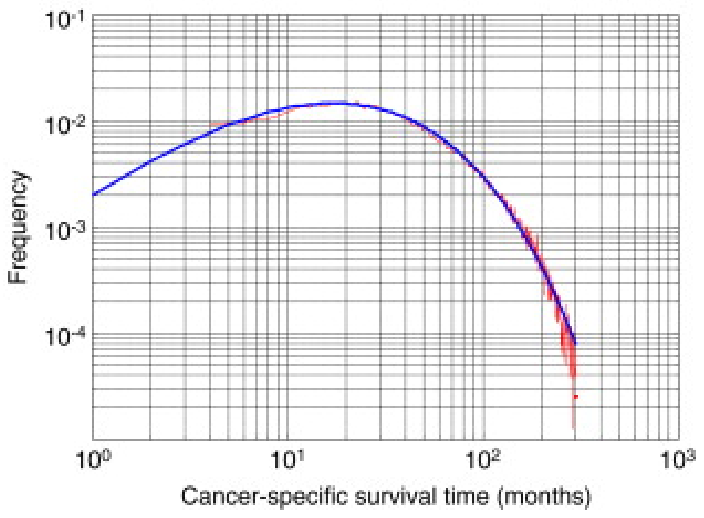}
\caption{Survival time statistics of breast cancer patients once
diagnosed with cancer $(t=0)$, both in a linear and double
logarithmic plot. Only patients that die from cancer are included
in the statistics. The solid line is the superstatistical model
prediction \cite{chen}. }
\end{figure}

\section{Maximum entropy principles, superstatistical path integrals,
and more}\label{sec6}

We finish this article by briefly mentioning some other recent
interesting theoretical developments.

One major theoretical concern is that a priori the
superstatistical distribution $f(\beta)$ can be anything. But
perhaps one should single out the really relevant distributions
$f(\beta)$ by a least biased guess, given some constraints on the
complex system under consideration. This program has been
developed in some recent papers \cite{abc,straeten}. There are
some ambiguities which constraints should be implemented, and
how. A very general formalism is presented in \cite{straeten},
which contains previous work \cite{abc,crooks, naudts}  as
special cases. The three important universality classes discussed
above, namely $\chi^2$ superstatistics, inverse $\chi^2$
superstatistics and lognormal superstatistics are contained as
special cases in the formalism of \cite{straeten}. In principle,
once a suitable generalized maximum entropy principle has been
formulated for superstatistical systems, one can proceed to a
generalized thermodynamic description, get a generalized equation
of state, and so on. There is still a lot of scope of future
research to develop the most suitable formalism. But the general
tendency seems to be to apply maximum entropy principles and
least biased guesses to nonequilibrium situations as well. In
fact, Jaynes \cite{jaynes} always thought this is possible.

Another interesting development is what one could call a
superstatistical path integral. These are just ordinary path
integrals but with an additional integration over a parameter
$\beta$ that make the Wiener process become something more
complicated, due to large-scale fluctuations of its diffusion
constant. Jizba et al. investigate under which conditions one
obtains a Markov process again \cite{jizba}. It seems some
distributions $f(\beta)$ are distinguished as making the
superstatistical process simpler than others, preserving
Markovian-like properties. These types of superstatistical path
integral processes have applications in finance, and possibly
also in quantum field theory and elementary particle physics.

In high energy physics, many of the power laws
observed for differential cross sections and energy spectra
in high energy scattering processes can also be explained using
superstatistical models \cite{wilknew, superscatter}.
The key point here is to extend the Hagedorn theory \cite{hage}
to a superstatistical one which properly takes into
account temperature fluctuations \cite{beck00,superscatter}.
Superstatistical techniques have also been recently used to describe the
space-time foam in string theory \cite{kings}.
\section{Summary}

{Superstatistics} (a 'statistics of a statistics') provides a
physical reason why more general types of Boltzmann factors (
e.g.\ $q$-exponentials or other functional forms) are relevant
for {\em nonequilibrium} systems with suitable fluctuations of an
intensive parameter. Let us summarize:

\begin{itemize}
\item
There is evidence for three major physically relevant {
universality classes}: $\chi^2$-superstatistics $=$ Tsallis
statistics, inverse $\chi^2$-superstatistics, and lognormal
superstatistics. These arise as {universal limit statistics} for
many different systems.

\item
Superstatistical techniques can be successfully applied to a {
variety of complex systems with time scale separation.}

\item The train delays on the British railway network are an example of
$\chi^2$ superstatistics = Tsallis statistics \cite{briggs}.

\item
A superstatistical model of {\em Lagrangian turbulence} \cite{prl}
is in excellent agreement with the experimental data for
probability densities, correlations between components, decay of
correlations, Lagrangian scaling exponents, etc. This is an
example of lognormal superstatistics \cite{prl}.

\item
Cancer survival statistics is described by inverse $\chi^2$
superstatistics \cite{chen}.

\item
The long-term aim is to find a good {thermodynamic description} for
general superstatistical systems. A generalized maximum entropy
principle may help to achieve this goal.

\end{itemize}

\end{document}